\documentstyle[epsf]{l-aa}
\newcommand{\diff}{{\rm d}}
\newcommand{\kcomp}{\hat{\kappa}_{\rm comp}}
\newcommand{\sdiff}{s_{\rm diff}}
\newcommand{\rcomp}{\rho_{\rm c}}
\newcommand{\ushock}{u_{\rm s}}

\newcommand{\gesim}{\,\raisebox{-0.4ex}{$\stackrel{>}{\scriptstyle\sim}$}\,}
\newcommand{\lesim}{\,\raisebox{-0.4ex}{$\stackrel{<}{\scriptstyle\sim}$}\,}
\newcommand{\tdecorrel}{t_{\rm d}}
\newcommand{\rgyro}{r_{\rm g}}

\newcommand{\lpar}{\lambda_{\scriptscriptstyle \|}}

\newcommand{\kpar}{\kappa_{\scriptscriptstyle \|}}
\newcommand{\lperp}{\lambda_{\scriptscriptstyle \bot}}
\newcommand{\kperp}{\kappa_{\scriptscriptstyle \bot}}
\newcommand{\kbohm}{\kappa_{\rm B}}
\newcommand{\bdiff}{D_{\rm M}}

\newcommand{\eqb}{\begin{eqnarray}}
\newcommand{\eqe}{\end{eqnarray}}

\begin{document}
\thesaurus{02.01.1; 02.04,2; 02.16.1; 02.19.1; 09.03.2; 09.19.2}
\title{Stochastic particle acceleration at shocks in the presence of 
braided magnetic fields}
\author{J.G. Kirk\inst{1}, P. Duffy\inst{1}, Y.A. Gallant\inst{2}}
\institute{Max-Planck-Institut f\"ur Kernphysik,
Postfach 10 39 80, D-69029 Heidelberg,
Germany
\and
Sterrenkundig Instituut Utrecht, 
Postbus 80000,
Utrecht 3508 TA, Netherlands}
\offprints{J.G. Kirk}
\date{Received 4 March 1996 Accepted \dots}
\maketitle
\markboth{J. G. Kirk et al.}{Particle acceleration in braided fields}
\begin{abstract}
The theory of diffusive acceleration of energetic particles at shock fronts
assumes charged particles undergo spatial diffusion in a uniform magnetic
field. If, however, the magnetic field is not uniform, but has a stochastic
or braided structure, the transport of charged particles across the 
average direction of the field is more complicated. Assuming quasi-linear
behaviour of the field lines, the particles undergo sub-diffusion 
on short time scales.
We derive the propagator for such motion, which differs from the Gaussian 
form relevant for diffusion, and apply it to a configuration
with a plane shock front whose normal is perpendicular to the average field 
direction. Expressions are given for the acceleration time as a function of 
the diffusion coefficient of the wandering magnetic field lines and the 
spatial diffusion coefficient of the charged 
particles parallel to the local field. 
In addition we calculate the spatial dependence of the particle
density in both the upstream and downstream plasmas. In contrast to the 
diffusive case, the density of particles at the shock front is lower than
it is far downstream. This is a 
consequence of the partial trapping of particles
by structures in the magnetic field. As a result, the spectrum of accelerated
particles is a power-law in momentum which is steeper than in the 
diffusive case. For a phase-space density $f\propto p^{-s}$, we find
$s=\sdiff[1+1/(2\rcomp)]$, where $\rcomp$ is the compression ratio of the 
shock front and $\sdiff$ is the standard result of diffusive acceleration:
$\sdiff=3\rcomp/(\rcomp-1)$. 
A strong shock in a monatomic ideal gas yields a spectrum of $s=4.5$. 
In the case of electrons, this
corresponds to a radio synchrotron spectral
index of $\alpha=0.75$.
\end{abstract}

\section {Introduction}
\label{intro}
Diffusive particle acceleration at shock fronts has been 
advanced as an explanation for many astrophysical phenomena involving
non thermal particles 
(for a review see Blandford \& Eichler~\cite{blandfordeichler87}).
A key to the success of this theory is its simplicity: particles are
assumed to diffuse in space upstream and downstream of a shock front,
which is treated as a simple discontinuity in the velocity of the
scattering centres. The well-known result is a power-law spectrum
of accelerated particles $f(p)\propto p^{-\sdiff}$ with an index
which is independent of the diffusion coefficient, being
determined solely by the compression ratio $\rcomp$ of the shock front:
$\sdiff=3\rcomp/(\rcomp-1)$.

In a recent {\em Letter} (Duffy et al.~\cite{duffyetal95})
it was pointed out that the transport of particles 
is more complicated in a 
\lq braided\rq\ magnetic field, i.e., one which has a random 
component causing field lines to wander around when projected onto
the plane
normal to their average direction. In particular, it was shown that
there may exist a regime of sub-diffusive transport, in which the average 
square displacement of a particle from its origin grows with the 
square-root of elapsed time. This differs sharply from
the linear growth expected
in the case of ordinary diffusion and reflects the restraining
influence or trapping (in a stochastic sense) of 
charged particles by the magnetic field. 
At shock fronts where particles must be transported across the direction of the
mean magnetic field in order to be accelerated, 
this leads to a modification of the acceleration rate.

In Sect.~\ref{transport} we present a detailed discussion of the
transport of particles in a braided field, pointing out the
diffusive and sub-diffusive regimes. There are 
two distinct physical effects which come into play here:
\lq microscopic\rq\ diffusion 
i.e., that which would pertain in a uniform 
magnetic field as a result of fluctuations with a length scale approximately
equal to the gyro radius of the particle concerned,
and \lq macroscopic\rq\ diffusion of the field lines 
themselves, which results from 
fluctuations of much longer length scale.
The importance of braiding, or \lq field line wandering\rq\ as it 
is called in the astrophysical literature
(Getmantsev~\cite{getmantsev63}, 
Jokipii \& Parker~\cite{jokipiiparker69},
Jokipii~\cite{jokipii73}) can be quantified in
terms of a single parameter $\Lambda$ related to the 
correlation lengths of the macroscopic magnetic field fluctuations and to the 
relative strength of the macroscopic and microscopic fluctuations. 
In the case of plasmas in which braiding is important, a further distinction
must be made concerning time scales. For long times, the transport is  
diffusive, and the spatial diffusion coefficient 
depends on a combination of the microscopic and macroscopic diffusivities.
On short time scales, the transport is sub-diffusive. 
In this case, we present the analytic form of 
the particle propagator, pointing out that it tends
to confine or trap the particles in a 
stochastic sense and is thus non-Markovian.

These ideas are applied in Sect.~\ref{shock}  
to find the distribution of energetic 
particles accelerated in the vicinity of a 
shock front whose normal is perpendicular
to the average direction of the field. At such 
shocks, the confining property of sub-diffusion
has a strong influence on the acceleration process,
since this is closely connected with the ability of 
particles to perform multiple crossings and recrossings
of the shock front before being advected away with
the downstream plasma. Assuming the angular distribution 
of the particles is kept close to isotropy by scattering off the 
microscopic fluctuations (an assumption which we discuss in 
Sect.~\ref{shock}), we compute
the spectrum and the 
spatial dependence of the density of accelerated particles,
using the exact form of the 
sub-diffusive propagator. Sub-diffusion manifests itself in a steeper 
spectral index than that given by the standard formula for diffusive 
acceleration. It also leads to a gradient of the particle density in the 
downstream plasma -- a quantity which is strictly uniform in the 
diffusive case.
The transition between sub-diffusive behaviour on short time scales and 
diffusive behaviour on longer ones can be investigated by constructing a 
propagator which switches from one form to the other at a certain point 
in time, which we call the decorrelation time. In Sect.~\ref{shock}  
results are found using such a propagator, and the value of the 
decorrelation time for which particle acceleration is strongly affected 
by sub-diffusion is deduced.

Finally, in Sect.~\ref{discussion}, we briefly discuss a few of the
implications and limitations of our results. In particular, we point out that 
in synchrotron sources in which field line wandering and acceleration 
of electrons at shock fronts is important, systematically steeper 
spectra should be observed from locations at which the shock is 
predominantly perpendicular.
\section{Cross-field transport}
\label{transport}
Our approach to the transport process involves separating it 
into two parts
(Chuvilgin \& Ptuskin~\cite{chuvilginptuskin93}). 
The first of these consists of 
macroscopic magnetic fluctuations of length scale
large compared to the gyro radius of the particle
concerned, 
which we 
characterise by a relative amplitude 
$b\equiv\left<|\delta B|\right>/\left<B\right>$.
As described in Duffy et al.~(\cite{duffyetal95}) we  assume these 
fluctuations lead to a quasi-linear type diffusion or wandering of the 
field lines in the plane perpendicular to the direction of the average 
field. This is described by a magnetic diffusivity $\bdiff$, which,
in terms of cartesian coordinates with the $z$-axis along the average 
field, is defined by:
\eqb
{\left<\Delta x^2\right>\over 2 s} &=& \bdiff
\enspace,
\eqe
where $\Delta x$ is the change in the $x$ co-ordinate upon 
travelling a distance $s$ along the field line. If the 
turbulence is characterised by correlation lengths $\lpar$ and
$\lperp$ along and across the average field, then assuming quasi-linear 
behaviour and adopting the normalisation of Kadomtsev 
\& Pogutse~(\cite{kadomtsevpogutse79}), we have  
\eqb
\bdiff&=&{b^2\lpar\over 4}
\enspace.
\eqe
(Note that different normalisations are used by 
Achterberg \& Ball~(\cite{achterbergball94}) and 
Isichenko~(\cite{isichenko91a})).

The second component of the transport concerns length scales 
comparable to that of the gyro radius.
We assume that any anisotropy in the distribution function results in 
the rapid growth of magnetic fluctuations (Alfv\'en waves) which, in 
their turn, scatter the particles so as to remove the anisotropy. This 
process we model in terms of two spatial diffusion coefficients
$\kpar$ and $\kperp$ responsible for transport along and across the 
local field direction, respectively. In keeping with other treatments
(e.g., Achterberg \& Ball~\cite{achterbergball94}, 
Jokipii~\cite{jokipii87}) we parameterise these coefficients in terms 
of the gyro-Bohm diffusion coefficient 
$\kbohm\equiv\gamma v^2 mc/(3eB)$, for 
a particle of mass $m$, charge $e$ moving at speed $v$, with
Lorentz factor $\gamma=(1-v^2/c^2)^{-1/2}$ in a magnetic field $B$:
\eqb
\kpar &=& {\kbohm\over\epsilon}
\nonumber\\
\kperp &=& {{\epsilon\kbohm}\over{(1+\epsilon)}}
\label{eqkperp}\enspace.
\eqe
Here $\epsilon\le1$ 
is the 
ratio of the energy density in microscopic fluctuations to that in the 
average magnetic field. In the 
picture in which the scattering is modelled by the 
\lq\lq$\tau$\rq\rq\ operator 
(Chuvilgin \& Ptuskin~\cite{chuvilginptuskin93}),
$\epsilon$ 
is the ratio of the collision rate to the gyro-frequency. 

Thus we have on the one hand microscopic fluctuations 
of relative amplitude $\sqrt{\epsilon}$ which are
responsible for the diffusion
of particles along and across the local field and, 
on the other, macroscopic fluctuations 
of relative amplitude $b$ which are 
responsible for diffusion of the field lines.
We shall assume $\epsilon\ll1$, which means 
that on the microscopic scale, particles are closely tied to field 
lines along which they diffuse. Transport across the local field 
direction is therefore severely restricted on the microscopic scale.

The important physical process in such a picture has been
described by Rechester and Rosenbluth~(\cite{rechesterrosenbluth78}):
because of the exponential divergence of neighbouring field lines,
a particle's orbit will take it out of regions of correlated magnetic
field. This can happen either because the particle diffuses a small 
distance across the field by scattering off the microscopic fluctuations,
or, in the case of very small $\epsilon$, because the finite size of 
its gyro orbit causes it to encounter uncorrelated field lines 
(Isichenko~\cite{isichenko91a}). The latter case we term 
\lq\lq ballistic\rq\rq\ propagation. It occurs
if the particle decorrelates from the field before
having a chance to diffuse microscopically, which translates into
the condition
\eqb
{b^2\over\epsilon}>{\lperp^2\over\lpar\rgyro}\log\left({\lperp\over\rgyro}\right)
\eqe
(Duffy et al.~\cite{duffyetal95}). 

Ballistic propagation has been considered by 
Achterberg \& Ball~(\cite{achterbergball94})
in connection with the radio emission of supernovae. 
The particles propagate diffusively,
with an effective spatial diffusion coefficient across the field given by
\eqb
\kperp^{\rm ballistic}&=& v\bdiff
\enspace,
\eqe  
and the standard theory of diffusive acceleration at shocks applies.
However, in this paper we will assume the self-excited microscopic turbulence 
is sufficiently strong to enable particles to scatter before they decorrelate from the 
magnetic field. In this case, one can define a dimensionless parameter
\eqb
\Lambda&=&{b^2\lpar\over\sqrt{2}\epsilon\lperp}
\eqe
such that for $\Lambda\lesim1$ the macroscopic braiding of the field is
irrelevant, whereas it dominates cross-field transport for $\Lambda\gg1$.
If braiding is irrelevant, there is no anomalous transport regime, and 
no modification of the diffusive acceleration picture, provided the distribution
can still be considered isotropic (cf.~Achterberg \& Ball~\cite{achterbergball94}). 

The most interesting parameter regime is that in which both braiding and microscopic
scattering are important, which occurs for the parameter range
\eqb
{\lperp\over\sqrt{2}\rgyro}\log\left({\lperp\over\rgyro}\right) >
\Lambda\gg1
\enspace.
\eqe
Two kinds of propagation are then possible: for times less than that needed 
to decorrelate from the field $\tdecorrel$, the particles undergo 
sub-diffusion, which is a combination of 
diffusion along a fixed field line, which itself diffuses.
The defining characteristic of sub-diffusion is that the mean square
deviation of a particle in the $x$ direction from its position at time $t=0$ is
not proportional to $t$ as in ordinary diffusion, but rather to $\sqrt{t}$. 

The propagator $P_{\rm sub}(x,t)$, which is the 
probability of finding a particle
in the interval ($x$,$x+\diff x$) at time $t$, 
given that it was at the origin $x=0$ at $t=0$, is
found by simply folding the 
two gaussian propagators appropriate to diffusion 
of a particle along the field (i.e., in $s$) and 
of the field in $x$:
\eqb
P_{\rm sub}(x,t)&=& {1\over 4\pi\sqrt{\bdiff\kpar t}}
\nonumber\\
&&\int_{-\infty}^{+\infty}
\diff s {1\over\sqrt{|s|}}
\exp\left(-{x^2\over4\bdiff |s|}-{s^2\over4\kpar t}\right)
\label{subdiffprop}
\enspace.
\eqe
(Rax \& White~\cite{raxwhite92}, Duffy et al.~\cite{duffyetal95}). 
It is straightforward to confirm that the mean square value of $x$ 
increases as $t^{1/2}$ with this propagator.
Equation~(\ref{subdiffprop}) 
describes the motion of an injected particle for times 
shorter than a decorrelation time, i.e.,
\eqb
t<\tdecorrel&=&{(\lperp\log\Lambda)^2\over2\kpar}
\label{decorrtime}
\enspace.
\eqe
Subsequently, the particle decorrelates and undergoes
\lq\lq compound diffusion\rq\rq, which is the collisional
transport regime discussed by Rechester \& 
Rosenbluth~(\cite{rechesterrosenbluth78}).
Here the propagation is diffusive in character,
with an effective diffusion
coefficient given by a combination of macro- and microscopic effects:
\eqb
\kappa_{\rm comp}
&\approx&\kperp\left(1+{\Lambda^2\over\log\Lambda}\right)
\label{kcompound}
\enspace
\eqe
The associated gaussian propagator is 
\eqb
P_{\rm comp}(x,t)&=& {1\over \sqrt{4\pi\kappa_{\rm comp}t}}
\exp\left[-{x^2\over 4\kappa_{\rm comp}t}\right]
\label{compprop}
\enspace.
\eqe

Although we have derived the sub-diffusive propagator on the basis of 
quasi-linear macroscopic turbulence of the magnetic field, it is a phenomenon
which probably has a much wider importance. This is demonstrated,  for example, 
by the numerical work of 
Rax \& White~(\cite{raxwhite92}), where sub-diffusion is
found using a simple Taylor-Chirikov mapping to 
model the macroscopic field turbulence.
Of course, the formulae we quote for quantities such as the decorrelation time
(Eq.~\ref{decorrtime}) are not generally applicable 
to other models of turbulence.
Furthermore, they apply only 
in the absence of significant particle drifts and only 
if the 
magnetic field can be considered static. 
Both drifts and time-dependent fields may 
in fact be responsible for decorrelating the particle from a field line 
(Isichenko~\cite{isichenko91b})
before this is achieved by either 
diffusion across the field or by the finite size of the 
gyro orbit. Nevertheless,
even though it does not appear possible 
to model the macroscopic turbulence in an
astrophysical source in such detail, 
we may nevertheless expect the qualitative 
picture of 
a sub-diffusive regime on short time scales 
followed by a diffusive one on longer
time scales to be a generic feature.
\section{Acceleration at quasi-perpendicular shocks}
\label{shock}
In a uniform magnetic field, the 
microscopic fluctuations enable particles to diffuse
in space. From the theory of diffusive acceleration
at parallel shocks, it is
well-known that the particle distribution at the shock front 
is almost isotropic, provided that the speed of the shock is
small compared to the speed of the individual particles.
Furthermore, if the magnetic field is not directed along the shock
normal, but is oblique, then it has been shown that in the absence of 
cross-field transport the distribution is almost isotropic, given that 
the speed of the point of intersection of the magnetic field and the 
shock front is small compared to the particle speed. If the magnetic 
field makes an angle $\phi_{\rm up}$ with the shock normal as
measured in the rest frame of the upstream plasma, this 
can be expressed as 
$\cos\phi_{\rm up}\gg \ushock/v$, where $\ushock$ is the speed of 
advance of the shock front into the upstream plasma and $v$ is the 
particle speed (Kirk \& Heavens~\cite{kirkheavens89}). For a slowly 
moving shock front, and for relativistic particles, this condition is 
fulfilled unless the magnetic field lies 
almost exactly in the plane of the shock. 
In a braided field, such points are relatively rare. A supernova
remnant, for example, has a shock front moving at a speed of roughly 
$3000\,{\rm km\,s^{-1}}$, so that for a relativistic electron
$\ushock/v\approx 1/100$ and the distribution remains close to 
isotropy unless the field is aligned to within about $1^\circ$ of the
shock plane. In the following, we assume that in a braided field, the 
microscopic fluctuations are able to keep the distribution close to 
isotropy at all points on the shock front and so neglect the effect of 
the small patches where the magnetic field is almost tangential to the 
shock surface. 

\subsection{Sub-diffusion}
Several properties of the accelerated particle
distribution can be found solely in terms of the single particle propagator
$P(x-x',t-t')$. In particular, given a source function $Q(x,t)$ of
injected particles the spatial density distribution 
$n(x,t)$ is
\eqb
n(x,t)=\int_{-\infty}^{+\infty}\diff x'
\int_{-\infty}^{t}\diff t' P(x-x',t-t')Q(x',t')
\label{propeqn}\enspace.
\eqe
Consider the case where particles are injected at a 
constant rate $Q_0$ for $t>t_0$ on
a plane boundary initially at $x=0$ which moves at constant speed $u$ in the 
positive $x$--direction; i.e. $Q(x,t)=Q_0\delta(x-ut)H(t-t_0)$.  
The time asymptotic 
steady state distribution, which is found by allowing $t_0\rightarrow-\infty$, is  
\eqb
n(y)&=&Q_0\int_0^{\infty}\diff  t P(y+ut,t)
\label{nasymp}
\eqe
where $y=x-ut$ is the distance from the plane boundary. 
Far downstream of the 
boundary, $y\rightarrow -\infty$, we have  $n(y)\rightarrow Q_0/u$. This can be seen by 
observing that far away from the boundary, 
the distribution should relax to a homogeneous state. Seen from the 
rest frame of the boundary, the flux of particles across any plane $y={\rm constant}$ is just $un(y)$, and this 
must equal the rate of injection at $y=0$. Alternatively, one may note that for a physically 
realistic propagator it 
is possible at any finite time $t$ to choose spatial boundaries beyond which there are a negligible number of 
particles. 
Denoting these boundaries by $\pm x_0$, we can write
\eqb
\int_{-x_0}^{x_0}\diff x P(x,t)=1;\quad
P(x,t)=0 {\rm \ for \ }|x|>x_0
\eqe
Then, substituting $x'=x/\alpha$ we have
\eqb
\alpha \int_{-x_0/\alpha}^{x_0/\alpha}\diff x'
P(\alpha x',t)=1
\eqe
or, equivalently, $P(\alpha x,t)\rightarrow \alpha^{-1}\delta(x)$, as $\alpha\rightarrow\infty$.
This property, together with Eq.~(\ref{nasymp}), 
implies $n(\infty)=0$ (far upstream) and $n(-\infty)=Q_0/u$ (far downstream).

On the boundary, however, the particle density depends on the precise form of the propagator.
Evaluating the integral in Eq.~(\ref{nasymp}) with the sub-diffusive propagator Eq.~(\ref{subdiffprop}), one 
finds $n(0)=2Q_0/3U$, so that the ratio of the density far downstream to that at the boundary is
\eqb
{n(-\infty)\over n(0)}&=& {3\over 2}
\label{nratio}
\eqe
In sub-diffusion, therefore, there is a spatial gradient downstream, which is absent in the case of diffusion. The 
scale length associated with this gradient can be obtained from the sub-diffusive propagator and is 
$\bdiff^{2/3}\kpar^{1/3}/U^{1/3}$.

As in the case of diffusive acceleration, these properties of the 
\lq\lq{\em sub-diffusion-advection}\rq\rq\ problem 
allow us to find the steady state phase space density $f_0(p)$ 
[$=n_0(p)/(4\pi p^2)$] of particles accelerated at a shock, provided we 
may assume that the shock itself does not affect the spatial transport
(see Sect.~\ref{discussion}).
Given that particles suffer no energy losses during propagation
and are accelerated only on crossing the shock front, conservation of particle
number in momentum space requires the advected flux of particles
with momentum $p$ far downstream of the shock to equal
the flux of particles across this momentum level at the shock (assuming no particles
are advected in from far upstream). For an almost isotropic distribution, this flux is simply related to the 
momentum derivative at the shock (e.g., Drury~\cite{drury87}, Kirk et 
al.~\cite{kirketal94}).
Writing the distribution function far downstream as $f_2(p)$,
and denoting by $u_1$ and $u_2$ the upstream and downstream flow speeds in the shock
one finds
\eqb
{\diff \over \diff p}\left({4\pi p^3\over 3}(u_1-u_2)f_0(p)\right)
+4\pi p^2u_2f_2(p)&=&0
\enspace,
\label{conserv}
\eqe
giving a power law solution $f_{0,2}\propto p^{-s}$ with 
\eqb
s&=&3\left[1+{f_2\over f_0(\rcomp-1)}\right]
\label{sfnratio}
\eqe
In the case of sub-diffusive propagation, 
$f_2/f_0=n(-\infty)/n(0)=3/2$, so that
\eqb
s&=&{3\rcomp\over\rcomp-1}\left(1+{1\over 2\rcomp}\right)
\label{subindex}
\eqe
For a strong shock, $\rcomp=4$, Eq.~(\ref{subindex}) gives $s=4.5$, which is steeper
than the spectrum produced by diffusive acceleration. 
Physically, this arises because particles 
are tied to individual field lines and are more effectively 
advected away from the shock than in the diffusive case.

This derivation may be re-cast slightly in terms of an {\it average}
probability per cycle, $P_{\rm esc}$, for a particle to escape once it enters
the downstream region (Bell~\cite{bell78}). In the steady 
state the ratio of the far downstream flux, $n(-\infty)|u_2|$, 
to the almost isotropic
flux of particles of speed $v$ crossing the shock, $n(0)v/4$, leads to 
$P_{\rm esc}=[n(-\infty)/n(0)](4|u_2|/v)$. Diffusion, for which the downstream gradient vanishes, leads to 
$P_{\rm esc}=4|u_2|/v$, whereas sub-diffusion gives $P_{\rm esc}=6|u_2|/v$. Thus, particles which are 
sub-diffusing have a higher 
probability of being advected away from the shock than those which are
diffusing. Assuming isotropy, the mean increase 
of a particle's momentum on crossing and recrossing the shock is 
$\langle\Delta p\rangle=4|u_1-u_2|p/3v$ from which the power  law  index of the phase 
space distribution can be calculated: $s=3+pP_{\rm esc}/\langle\Delta p\rangle$.
For sub-diffusion, $P_{\rm esc}=6|u_2|/v$, we obtain Eq.~(\ref{subindex}).

It is straightforward to generalise the sub-diffusive result
to any propagator of the form 
\eqb
P(x,t)={1\over t^{\beta/2}}\Phi\left({x\over t^{\beta/2}}\right)
\label{genprop}
\eqe
which corresponds to transport with 
$\left<\Delta x^2\right>\propto t^{\beta}$.\footnote{We are grateful to 
Prof.\ Luke Drury for pointing out this result to us} 
Using Eq.~(\ref{nasymp}), the normalisation property
$\int_{-\infty}^{\infty}\diff x \Phi(x)=1$, and assuming 
$\Phi(x)$ is an even function of $x$, one finds 
$n(-\infty)=n(0)(2-\beta)$, leading to a spectral index
\eqb
s&=&{3\rcomp\over\rcomp-1}\left(1+{1-\beta\over\rcomp}\right)
\label{generals}
\enspace.
\eqe

\subsection{Compound diffusion}
So far we have considered only cases in which it is a 
good approximation to use a single form of the 
propagator to describe the particle transport. However, we have 
argued that sub-diffusion applies only for times 
less than $\tdecorrel$, which is that needed for a particle to 
decorrelate from its original field line. Especially when calculating 
the stationary density profile, it is obvious that particles 
injected long ago should really be described by the 
appropriate diffusive propagator Eq.~(\ref{compprop}). 
There are several ways in which a propagator could be constructed with
such properties, and one could certainly envisage using these in 
a numerical simulation. However, in order keep an analytically tractable form,
we have been compelled to use a crude approximation.
This consists of artificially changing the behaviour of each particle at a time
$\tdecorrel$ after it has been injected. Sub-diffusion is assumed at $t<\tdecorrel$
and diffusion for all $t>\tdecorrel$:
The particle propagator can then be written as
\eqb
P(x,t)&=&\left\lbrace
\begin{array}{ll}
P_{\rm sub}(x,t)&{\rm for}\ t<\tdecorrel\\
\\
\multicolumn{2}{l}{
\int_{-\infty}^\infty\diff x' P_{\rm comp}(x-x',t-\tdecorrel)P_{\rm sub}(x',\tdecorrel)
}\\
&{\rm for}\ t>\tdecorrel
\end{array}\right.
\label{mixedprop}
\eqe
Although this propagator permits diffusion on long time scales, it 
does not allow for any subdiffusion 
at all at $t>\tdecorrel$. In reality, one 
would imagine that particles perform a sequence of subdiffusive
episodes with a mean duration equal to $\tdecorrel$, between which 
they decorrelate from the magnetic structures. 
In contrast to this, propagator~(\ref{mixedprop}) simply
releases all particles from the correlated field after a time 
$\tdecorrel$. We have also investigated an
alternative picture in which particles 
are released from a correlated patch once they have diffused 
along the magnetic field for more than a \lq\lq decorrelation 
length\rq\rq. This approach is closer
to the physical picture of propagation in a 
stationary magnetic field, as discussed by 
Duffy~et~al.~(\cite{duffyetal95}). It leads to 
results which differ little from those described below.

\begin{figure}
\epsfxsize=9 cm
\epsffile{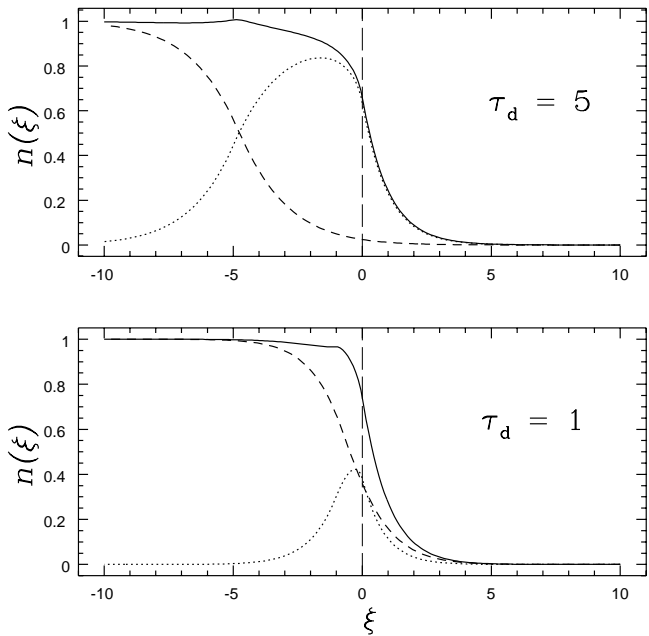}
\caption{\protect\label{profiles}
The particle density as a function of distance 
from a moving boundary for two values of the 
dimensionless decorrelation time $\tau_{\rm d}$. The dotted line represents 
$n_1(\protect\xi,\protect\tau_{\protect\rm d})$, i.e., those 
particles injected at the boundary 
within the last correlation time, which subsequently sub-diffused. 
The dashed line depicts $n_2(\protect\xi,\protect\tau_{\protect\rm d})$, 
i.e., particles injected 
before this. After an initial sub-diffusive period, these 
performed standard diffusion. 
The total particle density is indicated by the solid line. The length scale is in 
dimensionless units (see Eq.~\protect\ref{dimensionless}) 
and the density is normalised to unity far downstream.}
\end{figure}

Introducing dimensionless variables appropriate to the sub-diffusive case,
\eqb
\xi={u^{1/3}\over\bdiff^{2/3}\kpar^{1/3}}x\qquad
\tau={u^{4/3}\over\bdiff^{2/3}\kpar^{1/3}}t
\label{dimensionless}
\eqe
enables the dimensionless sub-diffusive and diffusive propagators to be written
\eqb
P_{\rm sub}(\xi,\tau)&=&{1\over2\pi}\int_0^\infty{\diff s\over \sqrt{s\tau}}\exp\left[
-\xi^2/(4s)-s^2/(4\tau)\right]
\nonumber\\
P_{\rm comp}(\xi,\tau)&=&{1\over\sqrt{4\pi\kcomp\tau}}
\exp\left[-\xi^2/(4\kcomp\tau)\right]
\eqe
Here $\kcomp=\kappa_{\rm comp}/(\bdiff^{2/3}\kpar^{1/3}u^{2/3})$ 
is the dimensionless form of the 
compound diffusion coefficient. 
This is closely related to the dimensionless decorrelation time
$\tau_{\rm d}$.
Evaluating the mean square deviation as a function of time, one finds
\begin{eqnarray}
\left<\Delta \xi^2\right>&=&\int_{-\infty}^\infty {\rm d}\xi\,
\xi^2 P(\xi,\tau)\\
&=&\left\lbrace
\begin{array}{ll}
4\sqrt{\tau/\pi}&{\rm for\ }\tau<\tau_{\rm d}\\
 \\
4\sqrt{\tau_{\rm d}/\pi}+2\kcomp(\tau-\tau_{\rm d})&{\rm for\ } \tau>\tau_{\rm d}
\end{array}
\right.
\end{eqnarray}
Since the propagator itself is continuous at $\tau=\tau_{\rm d}$, so too is
$\left<\Delta \xi^2\right>$. Choosing $\kcomp=1/\sqrt{\pi\tau_{\rm d}}$
means that the first derivative is also continuous, and this is the value we
adopt below. (It should be noted that the arguments leading, for example,
to Eq.~(\ref{kcompound}) determine $\kcomp$ only to within 
a factor of the order of unity.)

The stationary density at the shock for the mixed propagator 
Eq.~(\ref{mixedprop}) is made up of two parts: 
\eqb
\lefteqn{
n(\xi,\tau_{\rm d})=
\underbrace{\int_0^{\tau_{\rm d}}\diff\tau 
P_{\rm sub}(\xi+\tau,\tau)}_{n_1(\xi,\tau_{\rm d})}
}\hspace{-.5cm}
\nonumber\\
&+&
\underbrace{\int_{\tau_{\rm d}}^\infty\diff\tau
\int_{-\infty}^\infty\diff\xi' P_{\rm comp}(\xi+\tau-\xi',\tau-
\tau_{\rm d})  P_{\rm sub}(\xi',\tau_{\rm d})}_{n_2(\xi,\tau_{\rm d})}
\enspace.
\eqe
The density $n_1(\xi,\tau_{\rm d})$ consists of those particles 
which have been injected less than one 
decorrelation time ago, and have consequently propagated by 
pure sub-diffusion. Particles injected before this 
contribute to $n_2(\xi,\tau_{\rm d})$, and have undergone a period of 
Markovian diffusion after decorrelating 
from their original field line. Evaluating these quantities involves, 
for $n_1$, the numerical computation  of a 
double integral. On the other hand, the expression for $n_2$ may be 
reduced analytically to a single integral.
Figure~\ref{profiles} shows these quantities for two values of the 
dimensionless decorrelation time. One feature apparent from this figure is the 
irregular behaviour at a distance $\xi\approx\tau_{\rm d}$ into the downstream
medium. This is where, on average, our model propagator releases
particles into diffusive propagation. The precise form of the density profile 
at this point is sensitive to the value chosen for $\kcomp$. Since a realistic 
propagator would not decorrelate particles all at the same time, and
would also
reimpose sub-diffusion after decorrelation, the feature displayed 
has no direct physical significance. Nevertheless, this model 
propagator enables one
to get a qualitative picture of the density profile further upstream and downstream,
and also indicates the length scale on which a more realistic profile would 
vary.

\begin{figure}
\epsfxsize=9 cm
\epsffile{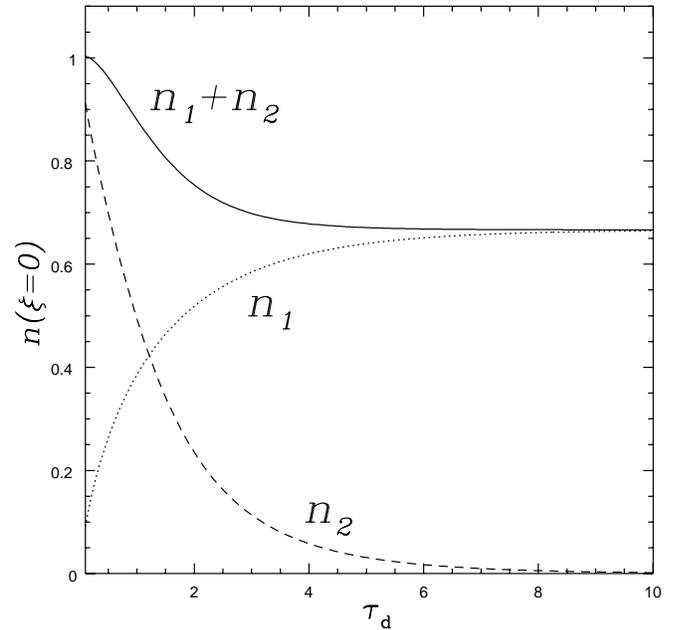}
\caption{\protect\label{shockden}
The density of particles at the shock front (i.e., moving boundary) as a
function of the decorrelation time. Shown are the recently 
injected, sub-diffusing particles
$n_1$, the older, diffusing particles $n_2$ and the total density.
As in Fig.~\protect\ref{profiles}, the density is normalised to 
take the value unity far
downstream.}
\end{figure}

\begin{figure}
\epsfxsize=9 cm
\epsffile{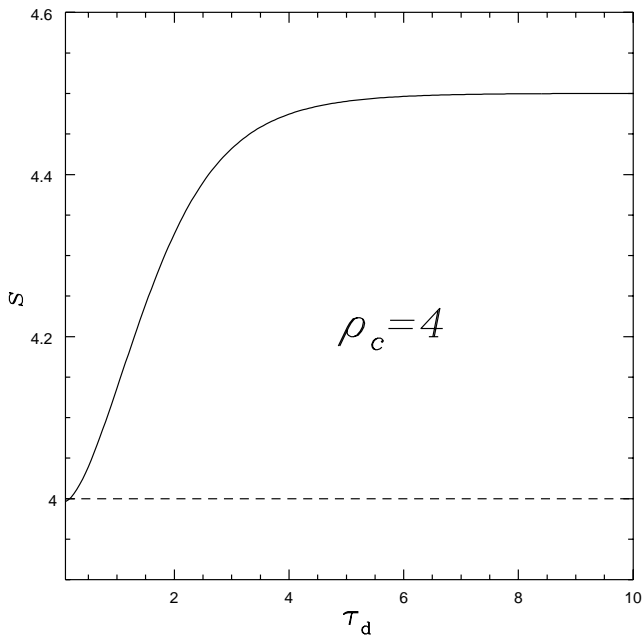}
\caption{\protect\label{sindex}
The power law index of accelerated particles at a shock front 
of compression ratio~4, computed according to 
Eq.~(\protect\ref{sfnratio}). It is assumed that after 
injection at the shock front particles propagate by sub-diffusion for times less than 
$\protect\tau_{\protect\rm d}$. 
Subsequently they diffuse. Only if $\protect\tau_{\protect\rm d}$ 
is short is the relatively flat spectrum of 
diffusive acceleration maintained. The time axis is in dimensionless units given in 
Eq.~(\protect\ref{dimensionless}).}
\end{figure}

The power-law index of accelerated particles can be found from the 
values of $n_1$ and $n_2$ at the shock 
front using Eq.~(\ref{sfnratio}). 
In this case, $n_1$ also reduces to a single numerical integration. 
These quantities are plotted as a function of $\tau_{\rm d}$ in Fig.~\ref{shockden}.
The corresponding power law index of accelerated particles is 
shown in Fig.~\ref{sindex}.
Here it can be seen that the spectrum steepens rapidly as the decorrelation time
increases. The index associated with sub-diffusion 
is achieved for decorrelation times  
greater than a few in dimensionless units, i.e., 
\eqb
\tdecorrel\gesim{\bdiff^{2/3}\kpar^{1/3} \over u_2^{4/3}}
\eqe
This quantity is approximately equal to the acceleration time found 
by Duffy et al.~(\cite{duffyetal95}). Thus, the steeper sub-diffusive
spectrum applies when particles are able to increase their energy 
substantially before decorrelating from the magnetic field.

\section{Discussion}
\label{discussion}

The main conclusion of this paper is that stochastic test 
particle acceleration at a strong shock which is predominantly perpendicular 
does not necessarily produce the
canonical $n(p)\propto p^{-2}$ spectrum expected from diffusive particle acceleration. 
Although much depends on the length scales and 
relative strengths of both the macroscopic inhomogeneities and the microscopic 
scattering, which, taken together, determine the decorrelation time $\tdecorrel$,
Eqs.~(\ref{subindex}) and (\ref{generals}) show that the 
spectrum produced depends in general on the 
properties of the turbulence, through the parameter $\beta$ where 
$\left<\Delta x^2\right>\propto t^{\beta}$. 

We have made several simplifying assumptions in order to arrive at an analytic
treatment of this problem.
As in the case of diffusive acceleration, we have assumed that the presence of the shock front does not have an 
important effect on the spatial transport of the particles, i.e., that the solution of the particle transport problem 
at a shock front can be obtained from a consideration of the case in which an imaginary boundary moves 
through a uniform stationary medium. This is certainly an idealisation. In a realistic situation the plasma 
upstream and downstream of a shock front probably supports turbulence of different character and amplitude. 
Furthermore, the plasma velocity and the magnetic field strength are discontinuous. Thus, we should allow for 
some change in the parameters of the propagator on crossing the shock, and perhaps also for a change of its 
functional form. Nevertheless, we feel that the essential physical feature introduced by the anomalous transport 
process -- the steepening of the spectrum -- is captured in our simple approach. This is because the nature of 
the transport in the upstream medium has very little effect on the particle spectrum, provided all particles 
entering this region are returned to the shock. The slope of the spectrum is determined by competition between 
energy gain on crossing the shock and escape downstream. In the case of sub-diffusion, a particle is always tied 
to the same field line. If we consider a particle at the shock front moving into the 
downstream medium, it is evident that its escape probability is determined by the 
average distance it can travel along its field line before encountering the shock again and does not depend on 
the nature of that portion of the field line which lies upstream. The situation is different if we are interested  in 
the spatial dependence of the particle density upstream, or in the time taken to perform a cycle of crossing and 
recrossing, since then the nature of propagation in the upstream field is important.

Another of our assumptions -- that the magnetic field lines themselves
diffuse -- is not readily lifted. Our calculation requires a specific model of the magnetic field, 
and we have too little knowledge of the properties of 
turbulence in the vicinity of shocks to do anything more realistic than simply assume diffusive 
behaviour. This corresponds to the case of 
sub-diffusion, $\beta=1/2$. However, 
any situation in which the magnetic field plays a role in 
inhibiting particle transport is likely to resemble that of 
sub-diffusion, or, more generally, the case $\beta<1$. 
In this connection, it is interesting to note that 
steep spectra have been found in numerical simulations of 
acceleration in random magnetic fields by 
Ballard \& Heavens~(\cite{ballardheavens92}).
Although there is no well-defined average field direction in their
computations, so that the shock cannot be described as quasi-perpendicular,
the fact
that particles diffuse freely in only one dimension, which does not always correspond
with the direction of the shock normal, leads one to suspect that 
here too, anomalous transport may be an important factor.
However, as well as the effects described above, their results may also be influenced 
by the relativistic flow speeds they assumed.

If particles are accelerated over several decades of momentum, 
as is thought to be the case in supernova 
remnants, it may be that different types of 
transport dominate in different ranges of $p$. 
One would expect that low energy particles which have relatively 
small gyro-radii compared to the correlation length of the magnetic field
might be more closely tied to 
the magnetic field lines than high energy ones, and so suffer more 
from sub-diffusive type effects.
We have not presented a detailed discussion of this effect here, 
which may, nevertheless, be important in the 
problem of the acceleration of cosmic rays (Duffy et al.~\cite{duffyetal95}).

Finally, the relative steepening of the spectral
index for $\beta=1/2$ also has important implications for the acceleration 
of electrons in astrophysical sources of synchrotron emission.
In particular, a correlation should arise between the obliquity of the shock and the 
spectral index of the radiation. At perpendicular shocks,
sub-diffusion can lead to particle spectra given by Eq.~(\ref{subindex}), 
which, even for uncooled electrons, would produce 
a relatively steep synchrotron spectrum of 
$F_{\nu}\propto \nu^{-0.75}$.

\noindent{\bf Acknowledgements:}
We are grateful to A.R.~Bell, R.O. Dendy and L.O'C.~Drury
for helpful and stimulating discussions.
This research was supported in part by the 
Commission of the European Communities
under Contract ERBCHRXCT940604.

\end{document}